\documentclass[prd,preprint,showpacs,preprintnumbers,amsmath,amssymb,eqsecnum,nofootinbib]{revtex4-1}

\usepackage{graphicx}
\usepackage{dcolumn}
\usepackage{bm}
\usepackage{subfigure}
\usepackage{color}

\begin{document}

\preprint{RUP-13-9}
\preprint{KEK-Cosmo-129}
\preprint{KEK-TH-1668}

\title{
Threshold of 
primordial black hole formation
}
\author{$^{1}$Tomohiro Harada}%
 \email{harada@rikkyo.ac.jp}
\author{$^{2}$Chul-Moon Yoo}
\author{$^{3,4}$Kazunori Kohri}
\affiliation{%
$^{1}$Department of Physics, Rikkyo University, Toshima, Tokyo 171-8501,
Japan} 
\affiliation{$^{2}$ Gravity and Particle Cosmology Group, Division of Particle and Astrophysical Science,
Graduate School of Science, Nagoya University, Furo-cho, Chikusa-ku, Nagoya 464-8602, Japan}
\affiliation{$^{3}$ Theory Center, Institute of Particle and Nuclear Studies,
KEK (High Energy Accelerator Research Organization),
1-1 Oho, Tsukuba 305-0801, Japan}
\affiliation{$^{4}$The Graduate University for Advanced Studies
(Sokendai), 1-1 Oho, Tsukuba
305-0801, Japan}
\date{\today}

\begin{abstract}
Based on a physical argument,
we derive a new analytic formula for 
the amplitude of density perturbation
at the threshold of primordial black hole formation in the 
Universe dominated by a perfect fluid with the equation of 
state $p=w\rho c^{2}$ for $w\ge 0$.
The formula gives $\delta^{\rm UH}_{H c}=\sin^{2}[\pi \sqrt{w}/(1+3w)]$
and $\tilde{\delta}_{c}=[3(1+w)/(5+3w)]\sin^{2}[\pi\sqrt{w}/(1+3w)]$, 
where $\delta^{\rm UH}_{H c}$ and $\tilde{\delta}_{c}$ are 
the amplitude of the density perturbation
at the horizon crossing time in the uniform Hubble slice
and the amplitude measure used in
 numerical simulations, respectively,
while the conventional one gives $\delta^{\rm UH}_{H c}=w$ and $\tilde{\delta}_{c}=3w(1+w)/(5+3w)$.
Our formula shows a much better agreement with the 
result of recent numerical simulations both qualitatively and 
quantitatively than the conventional formula.
For a radiation fluid, our formula gives 
$\delta^{\rm UH}_{H c}=\sin^{2}(\sqrt{3}\pi/6)\simeq 0.6203$ 
and 
$\tilde{\delta}_{c}=(2/3)\sin^{2}(\sqrt{3}\pi/6)\simeq 0.4135$.
We also discuss the maximum amplitude and the
cosmological implications of the present result.
\end{abstract}


\maketitle



\section{Introduction}
Primordial black holes may have formed from primordial fluctuations in
the early Universe~\cite{Zeldovich:1967,Hawking:1971ei}. Since
primordial black holes can in principle be observed at the present
epoch, current observations constrain the 
abundance of primordial black holes and thereby primordial 
fluctuations. In other words, primordial black holes can be 
used as a probe into the early Universe. 
This kind of analysis was first implemented by Carr~\cite{Carr:1975qj}. See 
Carr {\it et al.}~\cite{Carr:2009jm} for its latest update.
  
To constrain early Universe scenarios from the observational 
constraint of primordial black holes,
the formation threshold of the primordial black hole is very important. 
The conventional condition known as Carr's~\cite{Carr:1975qj} is that
a primordial black hole is formed if and only if    
the density perturbation $\delta_{H}$ when the fluctuation enters 
the horizon is in the range $w=\delta_{c} 
< \delta_{H}<\delta_{\rm max}= 1$,
where the equation of state $p=w\rho c^{2}$ is assumed. 
Although uncertainties in numerical factors of order unity in both 
the threshold and maximum values were noticed in the original paper,
the uncertainties have often been omitted 
in the subsequent literature.
However, the uncertainty of factor 2 in the threshold 
value $\delta_{c} $ results in enormous uncertainty in the prediction 
of the abundance of primordial black holes if we are given the 
power spectrum of the density perturbation because $\delta_{c}$
should be much greater than the standard deviation $\sigma$.
The maximum value $\delta_{\rm max}$, which 
was originally regarded as the separate universe
condition~\cite{Carr:1974nx}, has recently been shown 
~\cite{Kopp:2010sh,Carr:2013}
to be purely geometrical.

Since Nadezhin, Novikov, and Polnarev~\cite{Nadezhin:1978,Novikov:1980} 
pioneered the fully general relativistic 
numerical simulations of primordial black hole formation, 
the threshold of primordial black hole formation has been 
extensively investigated by numerical relativity~\cite{Niemeyer:1999ak,Shibata:1999zs,Musco:2004ak,Polnarev:2006aa,Musco:2008hv,Musco:2012au}.
Niemeyer and Jedamzik~\cite{Niemeyer:1999ak} reported the threshold value
$\delta_{c}\simeq 0.67-0.71$, which was later revised to the value $\simeq 0.43-0.47$ 
with a purely growing mode by Musco, Miller, and Rezzolla~\cite{Musco:2004ak}.
The latest value for a radiation fluid is given by 
$\delta_{c}\simeq 0.45-0.47$ and 
$\simeq 0.48-0.66$ depending on the parametrization of 
curvature profiles, as shown in Figs. 10 and 11 of 
Polnarev and Musco~\cite{Polnarev:2006aa}.
Moreover, Musco, and Miller~\cite{Musco:2012au} 
presented the numerical simulations of 
primordial black hole formation and the threshold values obtained 
for different values of $w$ in 
the range $0.01\le w\le 0.6$.

Khlopov and Polnarev~\cite{Khlopov:1980mg} pioneered the production of 
primordial black holes in the matter-dominated phase, 
where $w=0$, 
in the context of grand unification.
In the context of modern inflationary cosmology, 
the production of primordial black holes is interesting
not only in the radiation-dominated phase but also 
immediately after the inflationary phase, where $w\ll 1$
is effectively satisfied.
Suyama {\it et al.}~\cite{Suyama:2004mz,Suyama:2006sr} 
showed that primordial black holes
cannot be overproduced during the resonant preheating phase after the 
inflation but the production can be significantly enhanced 
in the universe undergoing tachyonic preheating. 
Alabidi {\it et al.}~\cite{Alabidi:2012ex,Alabidi:2013lya} 
discussed primordial black hole formation in the matter-dominated phase 
immediately after the inflation, where the formation efficiency 
may be enhanced by the softness of the equation of state but 
suppressed due to the effects of nonspherical collapse dynamics.

In the current paper, we derive a new analytic formula for the 
threshold of primordial black hole formation for general values of $w$
for $w \ge 0$ based on a physical argument.
For this purpose, we use a spherically symmetric model of a uniform 
overdensity surrounded by an underdense 
compensating layer in the flat Friedmann background.
Fixing a gauge problem, we
then see a very good agreement of our analytic formula with the numerical 
result by Musco and Miller~\cite{Musco:2012au} both qualitatively and
quantitatively. 

This paper is organized as follows.
In Sec.~II, we briefly summarize the original analysis of the 
condition for primordial black hole formation. 
In Sec.~III, we present our analytic model, derive 
a matter-independent maximum amplitude of the density perturbation
and discuss apparent horizons in this model.
In Sec.~IV, we analyze the threshold of primordial black hole
formation in the matter-dominated universe and in the 
universe dominated by a perfect fluid.
In the latter case, we derive a new analytic expression for the
threshold value.
In Sec.~V, we clarify the gauge problem and 
compare our analytic formula 
with the numerical result.
In Sec.~VI, we discuss the probability distribution of perturbations.
Section VII is devoted to summary.
We follow the metric signature $(-,+,+,+)$ and the abstract index
notation by Wald~\cite{Wald:1984rg}.

\section{Brief summary of the original analysis}
The original analysis by Carr~\cite{Carr:1975qj} is based 
on the physical argument that 
for an overdensity to form a primordial black hole,  
the size of the overdensity at the maximum expansion $R_{\rm max}$
should be larger than the Jeans radius $R_{J}$ 
(the Jeans criterion) but 
smaller than the particle horizon size $R_{\rm PH}$,
which is comparable with the curvature scale of the overdense region.
The maximum size was considered as necessary for the 
overdense region not to be separated from the rest of 
the universe~\cite{Carr:1974nx}.
This implies 
\begin{equation}
 R_{J}\alt R_{\rm max} \alt R_{\rm PH}.
\label{eq:RJRmaxRph}
\end{equation}
Note that the particle horizon size is given by $R_{\rm PH}\sim c/\sqrt{8\pi
G\rho_{\rm max}/3}$, while the Jeans radius is given 
by $R_{J}\sim \sqrt{w} R_{\rm PH}$, 
where  $\rho_{\rm max}$ is the density of the overdense region 
at the maximum expansion and 
the equation of state $p=w\rho c^{2}$ is assumed.
The condition (\ref{eq:RJRmaxRph}) 
implies that the density perturbation 
$\delta_{0}$ of mass scale $M$ at $t=t_{0}$
must satisfy
\begin{equation}
w\left(\frac{M}{M_{H_{0}}}\right)^{-2/3}\alt \delta_{0}\alt \left(\frac{M}{M_{H_{0}}}\right)^{-2/3},
\end{equation} 
where $M_{H_{0}}$ is the mass enclosed within 
the horizon at $t=t_{0}$.
This roughly gives
\begin{equation}
 w\simeq \delta_{c}\alt \delta_{H}\alt \delta_{\rm max}\simeq 1,
\label{eq:Carr's_condition}
\end{equation} 
where $\delta_{H}$ is the density perturbation at the horizon 
crossing
and $\delta_{c}$ and $\delta_{\rm max}$ denote the threshold value and 
the maximum value of $\delta_{H}$ for primordial black hole formation,
respectively.
This is often known as Carr's condition for 
primordial black hole formation. 
For a radiation fluid $w=1/3$, this gives the often quoted value 
$\delta_{c}\simeq 1/3$.
A more precise argument to derive 
this condition will be described later in this paper.

As Carr~\cite{Carr:1975qj} indicated, if the equation of state is
sufficiently soft, nonspherical effects play important roles 
rather than the Jeans criterion. 
Kopp, Hofmann, and Weller~\cite{Kopp:2010sh} pointed out that the maximum value $\delta_{\rm max}$
is not directly related to the separate universe
but to the geometry of the overdense region.

\section{Density perturbation model and the maximum amplitude}
\subsection{Three-zone model}

\begin{center}
\begin{figure}[htbp] 
 \includegraphics[width=0.5\textwidth]{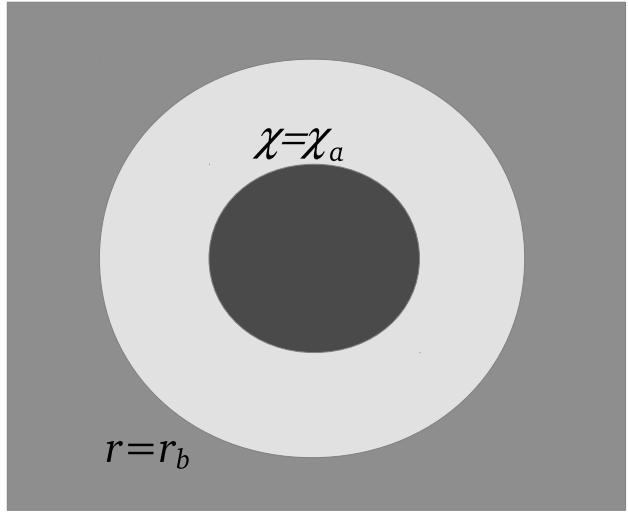}
\caption{\label{fg:density_perturbation_model} The schematic figure of
 the three-zone model of density perturbation.}
\end{figure}
\end{center}

Here we introduce a spherically symmetric model 
of density perturbation, which we will use for the analytic derivation
of the formation threshold and the maximum amplitude.
The model is schematically depicted in 
Fig.~\ref{fg:density_perturbation_model}.

The background universe is given by a flat Friedmann solution 
\begin{equation}
ds^{2}=-c^{2}dt^{2}+a_{b}^{2}(t)(dr^{2}+r^{2}d\Omega^{2}),
\end{equation}
where $d\Omega^{2}$ is the line element on the unit two-sphere.
The Friedmann equation is given by 
\begin{equation}
 \left(\frac{\dot{a_{b}}}{a_{b}}\right)^{2}=\frac{8\pi G\rho_{b}}{3},
\label{eq:flat_Friedmann_equation}
\end{equation}
where $\rho_{b}$ is the mass density of the background universe.
The overdense region is described by a closed Friedmann solution 
\begin{equation}
ds^{2}=-c^{2}dt^{2}+a^{2}(t)(d\chi^{2}+\sin^{2}\chi d\Omega^{2})
\end{equation}
or 
\begin{equation}
 ds^{2}=-c^{2}dt^{2}+a^{2}(t)\left(\frac{dr^{2}}{1-Kr^{2}}+r^{2}d\Omega^{2}\right),
\label{eq:curved_Friedmann_standard}
\end{equation} 
where $K=1$ and $r=\sin\chi$.
The Friedmann equation is given by 
\begin{equation}
 \left(\frac{\dot{a}}{a}\right)^{2}=\frac{8\pi
  G\rho}{3}-\frac{c^{2}}{a^{2}},
\label{eq:closed_Friedmann_equation}
\end{equation}
where $\rho$ is the mass density of the overdense region.
The overdensity is surrounded by an underdense layer which compensates 
the overdensity. We adopt a model where the overdense region is 
described by a closed Friedmann solution for $0<\chi<\chi_{a}$,
the surrounding underdense layer is matched with the overdense 
region at $\chi=\chi_{a}$, and the further surrounding 
flat Friedmann solution is matched with the compensating 
layer at $r=r_{b}$.
Thus, the areal radius of the overdense region is given by
$R_{a}=a\sin\chi_{a}$,
while that for the matching surface between the compensating layer
and the flat Friedmann universe is given by $R_{b}=a_{b}r_{b}$.
We call fluctuations with $0<\chi_{a}<\pi/2$ and $\pi/2<\chi_{a}<\pi$
types I and II, respectively, according to the notation of Kopp, 
Hofmann, and Weller~\cite{Kopp:2010sh}.
We note that the coordinates in Eq.~(\ref{eq:curved_Friedmann_standard})
cannot entirely cover the overdense region of type II fluctuation.

This model can be exact only for the dust case. In other cases,
inhomogeneity will penetrate the homogeneous regions through sound waves. 
To keep the model exact, we would need to introduce some unphysical 
matter field or shell in the compensating region. 
We here use this model, which can be called a 
``three-zone'' model, to obtain the threshold value of 
primordial black hole formation. 
This model can be justified at least for $w\ll 1$, where the effect
of pressure gradient force is very small.
It can also be justified at least in the early stage of evolution because
in the absence of decaying mode the inhomogeneity will be locally described by 
a homogeneous solution at each spatial point and the pressure 
gradient force can be neglected in accordance with the
Belinsky--Khalatnikov--Lifshitz conjecture~\cite{Belinsky:1970ew,Belinsky:1982pk}.

\subsection{Maximum amplitude of the density perturbation}
 
For convenience, we define the time-dependent 
density parameter $\Omega$ of the overdense region by 
\begin{equation}
 \Omega=\frac{8\pi
  G\rho}{3H^{2}}=
1+\frac{c^{2}}{a^{2}H^{2}},
\label{eq:Omega}
\end{equation}
where $H=\dot{a}/a$ is the Hubble parameter 
and we have used
Eq.~(\ref{eq:closed_Friedmann_equation}) in the last equality.
Defining the Hubble horizon radius $R_{H}=cH^{-1}$
in the overdense region, Eq.~(\ref{eq:Omega})
can be transformed to
\begin{equation}
 \left(\Omega-1\right)\left(\frac{R_{a}}{R_{H}}\right)^{2}=\sin^{2}\chi_{a}.
\label{eq:Omega-1_sin2chia}
\end{equation} 
This implies the left-hand side is time-independent and coincides with 
$\sin^{2}\chi_{a}$. 
The density perturbation $\delta$ of the overdense region to the background 
universe is defined by
\begin{equation}
 \delta=\frac{\rho-\rho_{b}}{\rho_{b}}.
\end{equation}
The density parameter $\Omega$ can then be related to $\delta$ according to
\begin{equation}
 \Omega=(1+\delta)\left(\frac{H_{b}}{H}\right)^{2},
\label{eq:Omega_and_delta}
\end{equation}
where $H_{b}=\dot{a}_{b}/a_{b}$ is the background Hubble parameter and
Eqs.~(\ref{eq:flat_Friedmann_equation}) and (\ref{eq:Omega}) are used.
It should be noted that the above relation is exact.
$\Omega$ is gauge-independent, while both
$\delta$ and $H_{b}$ are gauge-dependent.

The horizon crossing time is defined by the
equality between the areal radius of the overdense region, $R_{a}$, 
and the Hubble horizon of the background flat Friedmann universe,
$R_{H_{b}}=cH_{b}^{-1}$. Equations~(\ref{eq:Omega-1_sin2chia}) and
(\ref{eq:Omega_and_delta}) imply that the density perturbation
$\delta_{H}$ at the horizon crossing time is given by 
\begin{equation}
\delta_{H}=\left(\frac{H}{H_{b}}\right)^{2}-\cos^{2}\chi_{a},
\end{equation}
which trivially satisfies
\begin{equation}
\left(\frac{H}{H_{b}}\right)^{2}-1
< \delta_{H} \le \left(\frac{H}{H_{b}}\right)^{2}.
\label{eq:inequality_general}
\end{equation}
The maximum value is taken only for $\chi_{a}=\pi/2$, where the overdense 
region is a three-hemisphere. 
The lower limit corresponds to both $\chi_{a}=0$ and $\pi$, and the
latter corresponds to the separate universe limit of the overdense
region. 
The inequality (\ref{eq:inequality_general}) is automatically satisfied 
only if we assume the overdense region.
One value for $\delta_{H}$ generally
corresponds to two distinct configurations, the one of type I and the other of type II.
The maximum density does not correspond to 
the separate-universe configuration $\chi_{a}=\pi$,
as indicated by Kopp, Hofmann, and Weller~\cite{Kopp:2010sh}.

We can take a time slice on which the Hubble constants are the same
between the overdense and the background regions, i.e., $H=H_{b}$. 
We call this time slice the uniform Hubble slice. 
This is the case in the constant mean curvature slice, which is taken 
by Shibata and Sasaki~\cite{Shibata:1999zs}.
 
In the uniform Hubble slice, Eq.~(\ref{eq:Omega_and_delta}) 
implies that the time-dependent 
density parameter $\Omega$
and the density perturbation $\delta$ are directly related, i.e., 
\begin{equation}
 \Omega=1+\delta^{\rm UH},
\end{equation}
where $\delta^{\rm UH}$ denotes $\delta$ in the uniform Hubble slice.
Equation~(\ref{eq:Omega-1_sin2chia}) then implies
\begin{equation}
 \delta^{\rm UH}
\left(\frac{R_{a}}{R_{H}}\right)^{2}=\delta_{H}^{\rm UH}=\sin^{2}\chi_{a}.
\end{equation}
Therefore, $\delta^{\rm UH}(R_{a}/R_{H_{b}})^{2}$ is time-independent
and coincides with $\delta_{H}^{\rm UH}$. 
It immediately follows 
\begin{equation}
 0<\delta_{H}^{\rm UH}\le 1,
\end{equation}
where $\delta^{\rm UH}_{H}=1$
holds only for $\chi_{a}=\pi/2$. 
The above conclusion does not depend on  
the equations of state or even the matter fields.
The analysis does not invoke any
linearization with respect to the amplitude of the density perturbation.
It should be noted that we do not need to assume even 
the existence of maximum expansion here, 
although we will discuss it later in a different context.

If there is a maximum expansion phase of the overdense region, 
Eq.~(\ref{eq:closed_Friedmann_equation}) implies
\begin{equation}
 a_{\rm max}=\frac{c}{\sqrt{8\pi G \rho_{\rm max}/3}},
\end{equation}
where $\rho_{\rm max}$ is the density of the overdense region at the 
maximum expansion. In other words, $a_{\rm max}$ coincides with the Hubble 
horizon radius of the background flat 
Friedmann universe in the uniform density 
slice.  

\subsection{Trapped surfaces and apparent horizons}

In spherically symmetric spacetimes, we have a well-behaved quasilocal 
mass, which is called the Misner--Sharp mass~\cite{Misner:1964je}. 
The Misner--Sharp mass $M$ is defined as
\begin{equation}
 M=\frac{c^{2}}{2G}R\left(1-g^{ab}\nabla_{a}R\nabla_{b}R\right),
\end{equation}
where $R$ is the areal radius. 
This is closely related to the outgoing and ingoing null 
expansions, $\theta_{+}$ and $\theta_{-}$, respectively, 
where and hereafter 
we assume $\theta_{+}\ge \theta_{-}$ without loss of 
generality~\cite{Hayward:1994bu}.
If $2GM/(c^{2}R)>1$, we have $\theta_{+}\theta_{-}>0$.
A surface on which $\theta_{+}\theta_{-}>0$ is 
called a trapped surface.
A surface on which both $\theta_{+}$ and $\theta_{-}$ are negative
(positive) is said to be future (past) trapped.
If $2GM/(c^{2}R)<1$, we have $\theta_{+}\theta_{-}<0$.
A surface 
on which $\theta_{+}\theta_{-}<0$ is said to be untrapped. 
If $2GM/(c^{2}R)=1$, we have $\theta_{+}\theta_{-}=0$.
A surface on which $\theta_{+}\theta_{-}=0$ is 
called a marginal surface or an apparent horizon~\footnote{Strictly
speaking, the notion of an apparent horizon depends 
on the choice of a Cauchy surface on which it is defined. 
We here take the $t=$const. surface as a Cauchy surface.}.
A surface on which $\theta_{+}=0$ and $\theta_{-}<0$ ($\theta_{+}>0$ and
$\theta_{-}=0$ ) is called a future (past)
apparent horizon.
A future apparent horizon implies that no null geodesic 
congruence has positive expansion on it, 
which suggests the formation of a black hole. 
If the spacetime is asymptotically flat, the
existence of a future apparent horizon implies the existence of a future 
event horizon outside or coinciding with it~\cite{Hawking:1973uf}.
In fact, even if the spacetime is not asymptotically flat, a future apparent
horizon can be regarded as a black hole horizon. 
See Ref.~\cite{Hayward:1994bu} for more rigorous
terminology, definitions and proofs.

In the closed Friedmann spacetime, the areal radius and the 
Misner--Sharp mass are given by $R=a\sin\chi$
and 
\begin{equation}
 M=\frac{c^{2}}{2G}a\left[1+\left(\frac{\dot{a}}{c}\right)^{2}\right]
\sin^{3}\chi, 
\end{equation}
respectively. Since 
\begin{equation}
 \frac{2GM}{c^{2}R}=\left[1+\left(\frac{\dot{a}}{c}\right)^{2}\right]\sin^{2}\chi,
\label{eq:2MoverR}
\end{equation}
the apparent horizon, where $2GM/(c^{2}R)=1$, 
is given by a two-sphere
\begin{equation}
 \sin \chi=\left[1+\left(\frac{\dot{a}}{c}\right)^{2}\right]^{-1/2}.
\label{eq:apparent_horizon_trajectory}
\end{equation}
At the maximum expansion, there is a marginally trapped surface at 
$\chi=\pi/2$ or a great sphere. From Eqs.~(\ref{eq:2MoverR}) 
and (\ref{eq:apparent_horizon_trajectory}),
it follows that any
type II fluctuation immediately after the maximum expansion
necessarily has future trapped surfaces,
where $2GM/(c^{2}R)>1$, 
including $\chi=\pi/2$, and 
a future apparent horizon at $\chi\in (\pi/2,\chi_{a})$ which is given by 
Eq.~(\ref{eq:apparent_horizon_trajectory}).

\section{Threshold of primordial black hole formation}

\subsection{Matter-dominated universe}
In this section, we assume that the matter field is a dust, 
where our three-zone model is exact. 
The Friedmann equation for the overdense region is then given by 
\begin{equation}
 \dot{a}^{2}=\frac{A}{a}-c^{2},
\label{eq:closed_Friedmann_dust}
\end{equation}
where $A=8\pi G\rho_{0}a_{0}^{3}/3$ with $\rho=\rho_{0}$ and 
$a=a_{0}$ at $t=t_{0}$.
The solution of Eq.~(\ref{eq:closed_Friedmann_dust})
is given by 
\begin{eqnarray}
 a=\frac{a_{\rm max}}{2}(1-\cos\eta), \quad  t=\frac{t_{\rm
  max}}{\pi}(\eta-\sin\eta),
 \label{eq:scale_factor_solution_matter}
\end{eqnarray}
where $a_{\rm max}$ and $t_{\rm max}$ are given in terms of $a_{0}$ and
$\Omega_{0}$
as follows:
\begin{eqnarray}
 a_{\rm
  max}=\frac{\Omega_{0}}{\Omega_{0}-1}a_{0}=\frac{\Omega_{0}}{(\Omega_{0}-1)^{3/2}}cH_{0}^{-1},\quad
  t_{\rm max}=\frac{\pi}{2}\frac{a_{\rm max}}{c}, \label{eq:a_max_dust}
\end{eqnarray}
where Eq.~(\ref{eq:Omega}) is used.

The apparent horizon in the overdense region is given by 
\begin{equation}
 \eta=2\chi\quad \mbox{and}\quad \eta=2\pi-2\chi.
\end{equation}
If we concentrate on type I fluctuation, i.e., $0<\chi_{a}<\pi/2$,
the future apparent horizon corresponds to $\eta=2\pi-2\chi$.
Let us assume that a future apparent horizon exists when
the overdense region shrinks to $f$ $(0<f<1)$ 
times the maximum expansion, i.e., $a/a_{\rm max}=f$. 
Then, Eqs.~(\ref{eq:apparent_horizon_trajectory}) and
(\ref{eq:closed_Friedmann_dust}) 
yield
\begin{equation}
 \chi_{a}>\mbox{arcsin}\sqrt{f}.
\label{eq:nonspherical_effects}
\end{equation}
At the maximum expansion, the areal radius of 
the overdense region is given by
\begin{equation}
 R_{a,{\rm max}}=a_{\rm max}\sin\chi_{a}.
\label{eq:R_max}
\end{equation} 
This cannot be greater than $a_{\rm max}$.
The combination of Eqs.~(\ref{eq:nonspherical_effects}) and (\ref{eq:R_max})
means
\begin{equation}
 \sqrt{f}a_{\rm max}< R_{a,{\rm max}}\le  a_{\rm max}.
\label{eq:range1}
\end{equation}
Since we can rewrite $R_{a,{\rm max}}$ as 
\begin{equation}
 R_{a,{\rm max}}=a_{\rm max}\sin \chi_{a}=\frac{\Omega_{0}}{\Omega_{0}-1}a_{0}\sin \chi_{a}
 =\frac{\Omega_{0}}{\Omega_{0}-1}R_{a,0},
\end{equation}
where $R_{a,0}=a_{0}\sin \chi_{a}$ is the areal radius of the overdense 
region at $t=t_{0}$, using Eq.~(\ref{eq:a_max_dust}) we find 
\begin{equation}
f < (\Omega_{0}-1)\left(
\frac{R_{a,0}}{R_{H_{0}}}\right)^{2}\le 1,
\label{eq:exact}
\end{equation}
where $R_{H_{0}}=cH_{0}^{-1}$.
This is the condition for primordial black hole formation 
in terms of the quantities at $t=t_{0}$.
The above condition is exact, although
the factor $f$ is left unspecified.

It is a convention to express the condition for primordial black hole formation
in terms of the density perturbation $\delta_{H}$ at the horizon
crossing. 
As we have seen, 
$(\Omega_{0}-1)\left(R_{a,0}/R_{H_{0}}\right)^{2}$
is equal to the density perturbation $\delta_{H}^{\rm UH}$
at the moment of horizon crossing in the uniform Hubble slice.
Equation~(\ref{eq:exact}) can then be reduced to the condition 
in terms of $\delta^{\rm UH}_{H}$ as follows
\begin{equation}
 f < \delta_{H}^{\rm UH}\le 1.
\end{equation}
In the dust case, $f$ should be determined by 
considering 
the effects, such as 
caustics, inhomogeneity, and deviations from spherical symmetry
inside the overdense region; these effects can strongly affect
the collapse dynamics and 
then prevent the overdense region from becoming a black hole at the 
moment before the overdense region shrinks to $f$ times the maximum
expansion. These effects have been discussed by 
Khlopov and Polnarev~\cite{Khlopov:1980mg}.

\subsection{Universe dominated by a perfect fluid with $p=w\rho c^{2}$}
\label{subsec:threshold_derivation}

\subsubsection{Jeans radius and Carr's threshold}
We will see how the primordial black hole formation condition against the 
pressure gradient force is obtained with the three-zone model.
We assume the equation of state $p=wc^{2}\rho$ $(w>0)$. 
Except for $w\ll 1$, we can expect 
that the Jeans criterion gives the threshold of 
black hole formation
rather than the nonspherical effects.
For this case, the flat Friedmann solution is given by 
\begin{equation}
 a_{b}\propto t^{2/(3(1+w))}.
\label{eq:flat_Friedmann_solution}
\end{equation}
The Friedmann equation for the overdense region is given by 
\begin{equation}
 \dot{a}^{2}=Aa^{-(1+3w)}-c^{2},
\label{eq:closed_Friedmann_general}
\end{equation}
where $A$ is given by 
\begin{equation}
 A=\frac{8\pi}{3}G\rho_{0}a_{0}^{3(1+w)},
\label{eq:A_w}
\end{equation}
with $\rho=\rho_{a}$ and $a=a_{0}$ at $t=t_{0}$.

At the maximum expansion, the areal radius of 
the overdense region is given by
\begin{equation}
 R_{a,{\rm max}}=a_{\rm max}\sin\chi_{a}.
\end{equation} 
This cannot be greater than $a_{\rm max}$ due to spherical geometry, 
while this must be greater than
the Jeans radius $R_{J}$ of the overdense region at maximum expansion.
\begin{equation}
 R_{J}< R_{a,{\rm max}}\le  a_{\rm max}.
\label{eq:range2}
\end{equation}
The precise estimate of $R_{J}$ is not a trivial task. 
The standard Newtonian argument of the Jeans instability 
in a static and uniform gas cloud gives 
\begin{equation}
R_{J}=c_{s}\sqrt{\frac{\pi}{G\rho}},
\end{equation}
where $\rho$ and $c_{s}$ are the density and the sound speed
of the background uniform gas cloud, respectively. 
We may replace $c_{s}$ with
$\sqrt{w}c$ in the present case. 
Now we can adopt the following choice:
\begin{equation}
R_{J}=\sqrt{w}c\frac{1}{\sqrt{8\pi G\rho_{\rm max}/3}}=\sqrt{w}a_{\rm max}.
\label{eq:RJ_standard}
\end{equation}
Note that this is $\sqrt{w}$ times the Hubble radius 
of the background flat Friedmann universe 
in the uniform density slice.

Since Eqs.~(\ref{eq:Omega}), (\ref{eq:closed_Friedmann_general}) and 
(\ref{eq:A_w}) yield 
\begin{equation}
\frac{a_{\rm max}}{a_{0}}=\left(\frac{\Omega_{0}}{\Omega_{0}-1}\right)^{1/(1+3w)}
\end{equation}
and 
\begin{equation}
 a_{0}=\left(\Omega_{0}-1\right)^{-1/2}c H_{0}^{-1},
\end{equation}
Eq.~(\ref{eq:range2}) gives
the following exact relation:
\begin{equation}
w < (\Omega_{0}-1)\left(\frac{R_{a,0}}{R_{H_{0}}}\right)^{2}\le 1.
\end{equation}
Since
\begin{equation}
 (\Omega_{0}-1)\left(\frac{R_{a,0}}{R_{H_{0}}}\right)^{2}=\delta_{H}^{\rm UH}
\end{equation}
again, 
we find the following condition for primordial black hole formation:
\begin{equation}
 w< \delta^{\rm UH}_{H}\le 1.
\end{equation}
However, this is clearly dependent on the choice of $R_{J}$.
In other words, it is the choice of $R_{J}$
given by Eq.~(\ref{eq:RJ_standard}) that reproduces Carr's threshold.

\subsubsection{Refining the threshold}
It should be noted again that there is some ambiguity 
in the choice of the Jeans 
radius in Eq.~(\ref{eq:RJ_standard}) 
by a numerical factor of order unity.
Here we develop a physical argument to determine the numerical 
factor of the threshold value.

Defining the new variables $\tilde{a}$ and 
$\tilde{t}$~\cite{Harada:2004pe,Carr:2013} such that 
\begin{eqnarray}
 \tilde{a}=a^{1+3w}, \quad
 d\tilde{t}=(1+3w)\tilde{a}^{3w/(1+3w)}dt,
\label{eq:tilde_a_dtilde_t}
\end{eqnarray}
we can transform Eq.~(\ref{eq:closed_Friedmann_general}) into the dust
form:
\begin{equation}
 \left(\frac{d\tilde{a}}{d\tilde{t}}\right)^{2}=\frac{A}{\tilde{a}}-c^{2}.
\end{equation}
This can be integrated to give the parametric form of the solution,
\begin{eqnarray}
 \tilde{a}=\tilde{a}_{\rm max}\frac{1-\cos\eta}{2},\quad  
 \tilde{t}=\tilde{t}_{\rm max}\frac{\eta-\sin\eta}{\pi},
\label{eq:tilde_a_t}
\end{eqnarray}
where $\tilde{a}_{\rm max}$ and $\tilde{t}_{\rm max}$ are given as follows: 
\begin{eqnarray}
  \tilde{a}_{\rm max}= \frac{\Omega_{0}}{\Omega_{0}-1}\tilde{a}_{0},\quad 
\tilde{t}_{\rm max}=\frac{\pi}{2}\frac{\tilde{a}_{\rm max}}{c}.
\end{eqnarray}
Using the $(\eta,\chi)$ coordinates, the line element can be rewritten
in 
 the form
\begin{equation}
ds^{2}=\tilde{a}^{2/(1+3w)}\left[-\frac{1}{(1+3w)^{2}}d\eta^{2}
+d\chi^{2}+\sin^{2}\chi d\Omega^{2}\right].
\end{equation}
The apparent horizon in the overdense region is given by 
\begin{equation}
 \eta=2\chi\quad \mbox{and}\quad \eta=2\pi-2\chi.
\end{equation}
If we concentrate on type I perturbation, i.e., $0<\chi_{a}<\pi/2$,
the future apparent horizon corresponds to $\eta=2\pi-2\chi$.

The Jeans scale appears in the confrontation between 
the pressure gradient force and the gravitational force or 
equivalently between the sound crossing time and the free fall time. 
The sound wave propagates in the closed Friedmann geometry according to
\begin{equation}
 a\frac{d\chi}{dt}=\pm\sqrt{w}c.
\end{equation}
Using Eqs.~(\ref{eq:tilde_a_dtilde_t}) and (\ref{eq:tilde_a_t}),
this can be rewritten as
\begin{equation}
 \frac{d\chi}{d\eta}=\pm \frac{\sqrt{w}}{1+3w}
\label{eq:dchideta}
\end{equation}
in terms of $\eta$ and $\chi$. 
The solutions are given by 
\begin{equation}
 \eta=\pm \frac{1+3w}{\sqrt{w}}\chi+C_{\pm},
\end{equation}
where $C_{\pm}$ are constants of integration.

The rarefaction wave starts at the surface $\chi=\chi_{a}$ 
of the overdense region at $\eta=0$ and 
propagates inwardly to the center.
The compression wave also propagates from the center to the surface
outwardly, if there is any inhomogeneity within the overdensity. 
Since the region is initially expanding and the 
pressure gradient force generally pushes the fluid outwardly, 
if the sound wave crosses over the overdense region before 
the maximum expansion, the dynamics of the 
overdense region may be strongly affected
due to the pressure gradient force so that it may not reach the maximum
expansion but continue expanding. 
We can at least expect that the pressure gradient force significantly 
delays the collapse in this case.

\begin{center}
\begin{figure}[htbp]
 \includegraphics[width=0.7\textwidth]{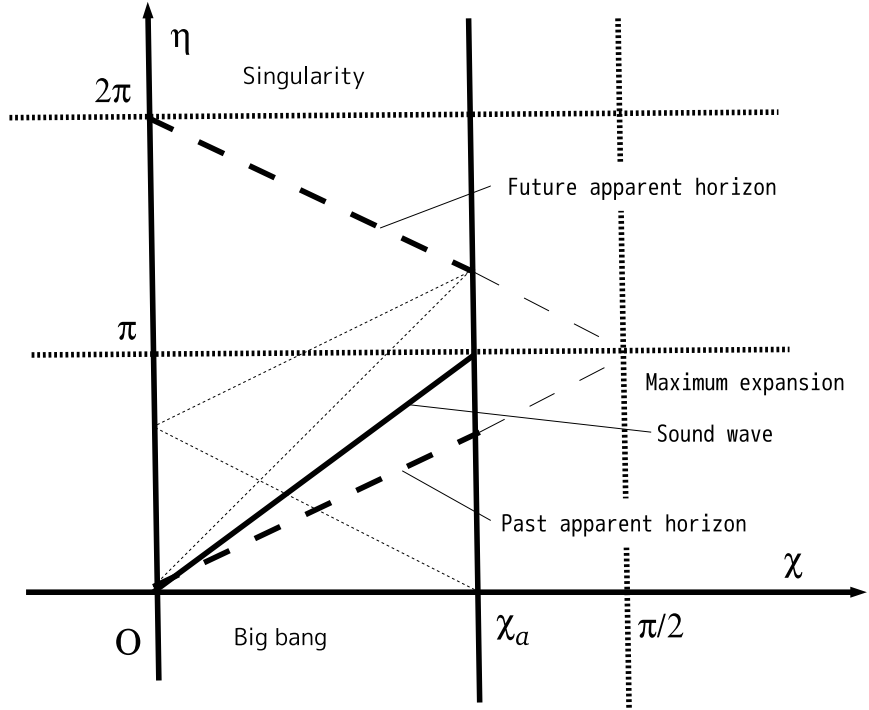}
\caption{\label{fg:eta_chi_diagram} 
The trajectories of 
the sound waves and apparent horizons in the $\eta\chi$ plane 
for the formation threshold. The sound wave just crosses over the 
radius of the overdense region from the big bang to the maximum 
expansion, which is denoted by a thick solid line. The 
stronger and weaker conditions are also shown by thin dashed lines.}
\end{figure}
\end{center}

This expectation motivates us to adopt the criterion that if and only 
if the sound wave crosses from the center to the surface outwardly 
or from the surface to the center inwardly 
before the maximum expansion, the pressure gradient force prevents 
the overdense region from becoming a black hole.
This requirement is naturally equivalent to the formation criterion that the 
sound crossing time over the radius be longer than the 
free fall time from the maximum expansion to complete collapse. 
See Fig.~\ref{fg:eta_chi_diagram}, which shows the trajectory of the 
sound wave for the threshold case, where 
the sound wave crosses over the radius of the overdense region 
at the same time of the maximum expansion.
The present criterion reduces to the following condition: 
\begin{equation}
 \chi_{a}>\frac{\pi\sqrt{w}}{1+3w}.
\end{equation}
This means that the Jeans scale $R_{J}$ 
at the maximum expansion can be identified with 
\begin{equation}
 R_{J}=a_{\rm max}\sin \left(\frac{\pi\sqrt{w}}{1+3w}\right).
\end{equation}
Therefore, we obtain the following formula for the threshold value of primordial black hole formation:
\begin{equation}
 \delta_{H c}^{\rm UH}=\sin^{2}\left(\frac{\pi\sqrt{w}}{1+3w}\right)
\label{eq:analytic_formula}
\end{equation}
and $\delta_{H}^{\rm UH}$ for primordial black hole formation 
must satisfy 
\begin{equation}
 \delta_{H c}^{\rm UH}< \delta_{H}^{\rm UH}\le 1.
\end{equation}
This can be considered as a (roughly) necessary and 
sufficient condition for primordial black
hole formation. 

Formula (\ref{eq:analytic_formula}) implies that 
$\delta_{H c}^{\rm UH}$ increases from 0, reaches a maximum 
value $\sin^{2}( \sqrt{3}\pi/6)\simeq 0.6203$ at $w=1/3$ and 
decreases to $1/2$, as $w$ increases from 0 to 1. 
$\delta_{H c}^{\rm UH}$ decreases as $w$ increases from $1/3$ 
because of the factor $1/(1+3w)$ on the 
right-hand side in Eq.~(\ref{eq:dchideta}).
This factor appears because the dynamical time of the collapse gets 
shortened by the contribution of the pressure to the
source of gravity.
$\delta_{H c}^{\rm UH}$ is approximated
as $\delta_{H c}^{\rm UH}\approx \pi^{2} w$ if $w\ll 1$, 
which is $\pi^{2}$ times the conventionally used 
Carr's threshold value $w$, 
and almost twice for a radiation fluid $w=1/3$.
This means that our analytic formula implies much less production 
efficiency for $w\ll 1$ and considerably less efficiency for $w=1/3$
than the conventional estimate. On the other hand, for $w\agt 0.6$, 
our formula gives a lower threshold value and hence implies higher 
production efficiency than the conventional estimate.

Although there are many other possible choices for the 
criterion of black hole formation, 
the present choice to derive Eq.~(\ref{eq:analytic_formula}) 
not only is physically natural 
but also shows a very good agreement
with the numerical result as we will see later.
To see this more explicitly, we further invent the following two 
conditions. The one is a stronger formation condition that the 
future apparent horizon must form before the 
sound wave crosses over the radius. This leads to 
\begin{equation}
 \chi_{a}>\frac{2\pi\sqrt{w}}{1+2\sqrt{w}+3w}
\quad 
\mbox{or} 
\quad 
 \delta_{H c}^{\rm UH}=\sin^{2}\left(\frac{2\pi\sqrt{w}}{1+2\sqrt{w}+3w}\right).
\end{equation}
The other is a weaker condition that the future apparent horizon 
must form before the sound wave propagates inwardly from the surface to 
the center and then outwardly back from the center to the surface. 
This leads to 
\begin{equation}
 \chi_{a}>\frac{\pi\sqrt{w}}{1+\sqrt{w}+3w}
\quad 
\mbox{or} 
\quad 
 \delta_{H c}^{\rm UH}=\sin^{2}\left(\frac{\pi\sqrt{w}}{1+\sqrt{w}+3w}\right).
\end{equation}

\section{Comparison with the numerical result}

\subsection{Density perturbation in the comoving slice}

We here study the density perturbation in the comoving slice.
For this purpose, we need to systematically introduce inhomogeneity, 
which arises from the big bang universe.
Indeed, 
Polnarev and Musco~\cite{Polnarev:2006aa} introduce a
time-independent function of $r$, $K=K(r)$, into the 
Friedmann--Robertson--Walker metric
(\ref{eq:curved_Friedmann_standard}) and obtain an asymptotic solution 
of the Einstein equation in the limit $t\to 0$, 
where all the hydrodynamical quantities
are nearly homogeneous with their perturbations 
being small deviations with the small parameter 
$\epsilon=(R_{H_{b}}/R_{a})^{2}$ but the curvature perturbations
can be arbitrarily large.
They call such solutions asymptotic quasihomogeneous solutions.
They explicitly obtain the first-order
solution in terms of $\epsilon$, which 
is consistent with a pure growing mode of superhorizon scale 
in the linear perturbation theory.
(See Ref.~\cite{Polnarev:2012bi} for higher-order solutions.) 
They use the first-order solution as initial data to simulate
the subsequent nonlinear evolution.

We here give the relationship between the density 
perturbations in the uniform Hubble slice and in the 
comoving slice.
The combination of Eqs. (32), (41), (57), and (85) of
Polnarev and Musco~\cite{Polnarev:2006aa} gives 
the first-order solution 
for the density perturbation $\delta$ in the following form:
\begin{equation}
 \delta^{\rm COM}_{1}=\frac{3(1+w)}{5+3w}K(r_{0})r_{0}^{2}\left(\frac{R_{H_{b}}}{R_{a}}\right)^{2},
\end{equation} 
where $\delta^{\rm COM}_{1}$ and $r_{0}$ 
denotes the first-order solution for the density perturbation
in the comoving slice and 
the comoving radius of the overdense region, respectively.
For the overdense region in our three-zone model, we have $K(r)=1$ and 
$r_{0}=\sin\chi_{a}$, and therefore
\begin{equation}
 K(r_{0}) r_{0}^{2}=\sin^{2}\chi_{a}=\delta_{H}^{\rm UH}.
\end{equation}  
Defining $\tilde{\delta}$ by 
\begin{equation}
 \tilde{\delta}=\delta^{\rm COM}_{1}\left(\frac{R_{a}}{R_{H_{b}}}\right)^{2},
\end{equation}
we find that this is time-independent and 
\begin{equation}
 \tilde{\delta}
=\frac{3(1+w)}{5+3w}\delta_{H}^{\rm UH}.
\label{eq:tilde_delta_COM_delta_H_UH}
\end{equation}
$\tilde{\delta}$ is used as the measure of the density
perturbation in the numerical simulations in 
Refs.~\cite{Musco:2004ak,Polnarev:2006aa,Musco:2008hv,Musco:2012au}.
Note that although $\tilde{\delta}$ is defined in terms of 
the first-order solution of the asymptotic quasihomogeneous solution,
the relation~(\ref{eq:tilde_delta_COM_delta_H_UH}) between 
$\tilde{\delta}_{c}$ and $\delta_{H}^{\rm UH}$ is exact.

\subsection{Comparison with the numerical result}

\begin{center}
\begin{figure}[htbp]
 \includegraphics[width=0.8\textwidth]{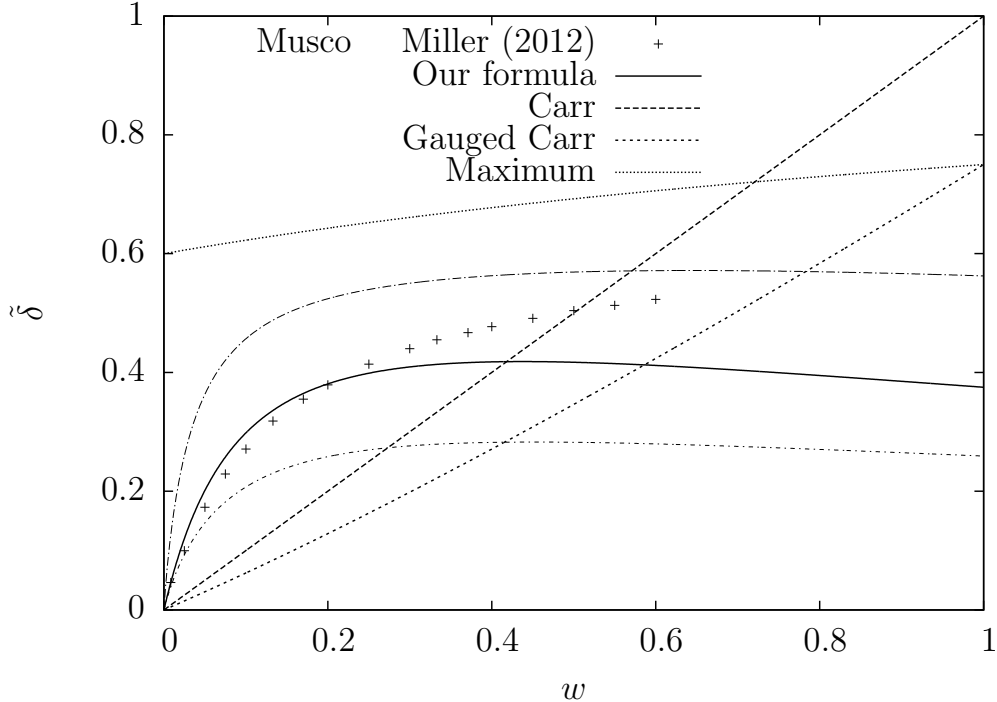}
\caption{\label{fg:comparison} 
The threshold values and the maximum value of the density
 perturbation variable $\tilde{\delta}$ 
in the comoving slice for different values of $w$. 
The crosses plot the result of numerical simulations by
Musco and Miller~\cite{Musco:2012au} 
for the profile parameter $\alpha=0$ or a Gaussian curvature profile.
The solid, long-dashed and dashed lines denote
 the analytic formula obtained in
 Sec.~\ref{subsec:threshold_derivation}, 
Carr's original formula and its gauged version, respectively. 
We also plot our stronger and weaker conditions with
thin dotted-dashed lines, which are discussed in
 Sec.~\ref{subsec:threshold_derivation}.
The short-dashed line denotes the 
geometrical maximum value, corresponding to a 
three-hemisphere.}
\end{figure}
\end{center}

The latest accurate estimate of the threshold value based on 
fully general relativistic numerical simulations has been given 
by Musco and Miller~\cite{Musco:2012au} for $0.01\le w \le 0.6$. 
Figure~\ref{fg:comparison} shows the comparison of our analytic formula 
with the numerical result shown in 
Fig. 8 of Ref.~\cite{Musco:2012au}. 
Since the numerical result is not so sensitive to the 
parameter $\alpha$ of the curvature profile function 
adopted in Ref.~\cite{Musco:2012au},  
we only plot the numerical result for $\alpha=0$ or a Gaussian profile 
for clarity.
According to Musco and Miller~\cite{Musco:2012au}, we here 
present the comparison with the perturbation variable 
in the comoving 
slice, $\tilde{\delta}$, which is directly related 
to the exact density perturbation in the uniform Hubble slice 
at the moment of horizon crossing, $\delta_{H}^{\rm UH}$, by 
Eq.~(\ref{eq:tilde_delta_COM_delta_H_UH}). 
In terms of $\tilde{\delta}$, our analytic formula gives
\begin{equation}
 \frac{3(1+w)}{5+3w}\sin^{2}\left(\frac{\pi\sqrt{w}}{1+3w}\right)=\tilde{\delta}_{c}
<\tilde{\delta}
\le \tilde{\delta}_{\rm max}=\frac{3(1+w)}{5+3w}.
\end{equation}

In Fig.~\ref{fg:comparison}, we plot our analytic formula
for the threshold $\tilde{\delta}_{c}$
together with Carr's original value
$w$ and its gauged value $3(1+w)w/(5+3w)$.
We also plot our stronger and weaker conditions in the same figure.
As we can see in Fig.~\ref{fg:comparison}, our analytic formula
agrees with the result of the numerical 
simulations within 20 \% approximately for $0.01\le w\le 0.6$.
Note that our analytic formula gives $\tilde{\delta}_{c}\approx 
3\pi^{2}w/5$ for $w\ll 1$, $(2/3)\sin^{2}(\sqrt{3}\pi/6)\simeq 0.4135$ 
for $w=1/3$ and 3/8 for $w=1$.
We also find that the numerical result can be qualitatively 
explained by our sinusoidal function rather 
than the straight line.
For larger values of $w$ ($w\agt 1/3$), 
our formula appears to systematically underestimate the threshold value.
For a radiation fluid ($w=1/3$), our formula gives a 
value smaller than the numerical result of Musco and Miller~\cite{Musco:2012au}
by 10 \% approximately.
However, we should note that 
the numerical result also should have dependence on 
the density profile. 
It has been reported~\cite{Polnarev:2006aa} 
that the threshold value for a radiation fluid is 
$\tilde{\delta}_{c}\simeq 0.45-0.47$ and 
$\simeq 0.48-0.66$ depending on the 
parametrization of curvature profiles as shown in 
Figs. 10 and 11 of Ref.~\cite{Polnarev:2006aa}. 
This suggests that the 20 \% deviation cannot be avoided within 
our simplified analytic model.  
Our formula shows a much better agreement for smaller values of $w$
than Carr's original formula and 
its gauged version, as expected.
Even for larger values of $w$, we can still see 
that our formula generally shows a better 
agreement both qualitatively and quantitatively
than the gauged version of Carr's formula.
We can also see that the numerical result of Musco and 
Miller~\cite{Musco:2012au} is between our stronger 
and weaker conditions.

Our threshold formula implies that the threshold values 
are approximately given by  
$\delta_{H c}^{\rm UH}\simeq 0.5-0.6$ and $\tilde{\delta}_{c}\simeq 0.4$
and for $ 1/3\alt w \alt 1$
and are not so sensitive to $w$ in this range. 
Our formula also suggests that primordial black holes
can be formed from type I fluctuations 
even for very hard equations of state, i.e., $w\simeq  1$,
because $\tilde{\delta}_{c}$ 
is well below $\tilde{\delta}_{\rm max}$.

\section{Probability distribution}

Conventionally, it has been assumed that 
the probability distribution for the density perturbation 
follows a Gaussian distribution. Then, the fraction $\beta_{0}(M)$
of the Universe
which goes into primordial black holes of mass scale $M$
at the formation epoch is given by 
\begin{eqnarray}
\beta_{0}(M)&=&\int_{\delta_{c}(M)}^{\delta_{\rm max}(M)}\frac{2}{\sqrt{2\pi\sigma^{2}(M)}}
\exp\left(-\frac{\delta^{2}}{2\sigma^{2}(M)}\right)d\delta \nonumber \\
&\simeq & \mbox{erfc}\left(\frac{\delta_{c}(M)}{\sqrt{2}\sigma(M)}\right) 
\nonumber \\
&\simeq& \frac{\sqrt{2}}{\sqrt{\pi}}\frac{\sigma(M)}{\delta_{c}(M)}
\exp\left(-\frac{\delta_{c}^{2}(M)}{2\sigma^{2}(M)}\right),
\end{eqnarray}
where $\sigma(M)$ is the standard deviation of the density perturbation
of mass scale $M$, the factor 2 comes from the Press--Schechter theory,
$\mbox{erfc}(x)$ is the complementary error function and 
we have assumed $\delta_{\rm max}\gg \delta_{c}\gg \sigma(M)$ 
in the second and last equalities. The last expression
is a consequence of the asymptotic expansion 
of $\mbox{erfc}(x)$ for $x\gg 1$.
In the above expression, $M$ just denotes the mass contained within the 
overdense region and may be different from the final black hole 
mass because of possible critical behavior~\cite{Niemeyer:1997mt} or mass accretion. 

However, as we have seen, the density perturbation $\delta$ 
has a finite maximum value and one value of $\delta$ generally 
corresponds to two perturbation configurations, $\chi_{a}$, 
the one of type I and the other of type II. The type II 
fluctuation is nonlinearly large, although $\delta$ may be very small.
The Gaussian assumption to $\delta$ implies the following unreasonable
consequence: 
a linearly small perturbation 
$\chi_{a}\simeq 0$, where the overdense region 
is only slightly bent, and a highly nonlinear perturbation 
$\chi_{a}\simeq \pi$, which is nearly separate 
from the rest of the universe,
would be realized with the same probability.

Recently, Kopp, Hofmann, and Weller~\cite{Kopp:2010sh} 
suggested that the curvature fluctuation is more suitable for 
the assumption of probability distribution.
The curvature fluctuation $\zeta$ is defined by the conformal factor of 
the three metric in the conformally flat form:
\begin{equation}
 ds_{3}^{2}=b^{2}(t)e^{2\zeta(t,s)}(ds^{2}+s^{2}d\Omega^{2}).
\end{equation}
The averaged curvature fluctuation $\bar{\zeta}$
is defined in Ref.~\cite{Kopp:2010sh} in terms of $\chi_{a}$ as
\begin{equation}
 \bar{\zeta}=\frac{1}{3}\ln 
\frac{3(\chi_{a}-\sin\chi_{a}\cos\chi_{a})}{2\sin^{3}\chi_{a}},
\label{eq:barzeta_chia}
\end{equation}
where $b(t)$ is chosen to be common between the overdense region 
and the background flat Friedmann region.
On the other hand, the peak value of the original variable 
$\zeta(t,0)$, which will be denoted just by $\zeta$, can be 
approximately expressed as~\cite{Kopp:2010sh} 
\begin{equation}
 \zeta\simeq -2\ln \cos\frac{\chi_{a}}{2}
\label{eq:zeta_chia}
\end{equation} 
in the present model, if the contribution from the compensating layer is negligible.
$\bar{\zeta}$ and $\zeta$ are plotted as functions of $\chi_{a}$ in
Fig. 4 of Ref.~\cite{Kopp:2010sh}.
$\zeta$ (or $\bar{\zeta}$) can be arbitrarily large
even for $R_{H_{b}}/R_{a}\ll 1$, where 
the density perturbation $\delta$ is sufficiently small. 
Moreover, unlike $\delta$,
$\zeta$ monotonically increases from 0 to $\infty$ as
$\chi_{a}$ increases from 0 to $\pi$. 
The threshold value can be derived by substituting 
$\chi_{a}=\arcsin\sqrt{\delta_{H c}^{\rm UH}}$ into the right-hand side of 
Eq.~(\ref{eq:zeta_chia}). Since
any type II fluctuation necessarily has a future apparent horizon
immediately after the maximum expansion, 
the threshold configuration must be of
type I and hence $0<\chi_{a}<\pi/2$.

We should note that $\zeta$ takes a value between $0$ and $\infty$, 
that it has one-to-one correspondence with 
the overdensity configuration $\chi_{a}$, 
and that $\zeta$ is proportional to $\delta$ in the linear regime.
For the above three facts, we can naturally extend 
a Gaussian distribution for $\zeta$ 
(or any other similar curvature variable) to the nonlinear regime, although 
this needs further justification. 
As a consequence of this assumption, a linearly small perturbation, 
i.e., $\chi_{a}\simeq 0$, is realized with high probability,
while a nearly separate universe, 
i.e., $\chi_{a}\simeq \pi$, is realized with extremely low probability.
That is, we have 
\begin{eqnarray}
 \beta_{0}(M)&=&\int_{\zeta_{c}(k_{\rm BH})}^{\infty}\frac{2}{\sqrt{2\pi P_{\zeta}(k_{\rm BH})}}
\exp\left(-\frac{\zeta^{2}}{2 P_{\zeta}(k_{\rm BH})}\right)d\zeta \nonumber \\
&= & \mbox{erfc}\left(\frac{\zeta_{c}(k_{\rm BH})}{\sqrt{2 P_{\zeta}(k_{\rm BH})}}\right) \nonumber \\
&\simeq & \frac{\sqrt{2P_{\zeta}(k_{\rm BH})}}{\sqrt{\pi}\zeta_{c}(k_{\rm BH})}\exp\left(-\frac{\zeta_{c}^{2}(k_{\rm BH})}{2 P_{\zeta}(k_{\rm BH})}\right),
\end{eqnarray}
where $P_{\zeta}(k)$ is the power spectrum of $\zeta$, 
$k_{\rm BH}=a_{b}H_{b}$ at the horizon crossing, 
and only in the last expression 
$\zeta_{c}^{2}(k_{\rm BH}) \gg P_{\zeta}(k_{\rm BH})$ is assumed.
Since $\zeta^{2}_{c}\gg P_{\zeta}(k_{\rm BH})$ is usually 
assumed, it is clear that the precise estimate of 
the threshold value $\zeta_{c}$ is very important.

For type I fluctuations, we find 
$
0<\bar{\zeta}<\bar{\zeta}_{h}=(1/3)\ln (3\pi/4)\simeq 0.2857,
$
where $\bar{\zeta}_{h}$ is the value for $\chi_{a}=\pi/2$.
Our analytic formula gives an expression for
$\bar{\zeta}_{c}$ as follows:
\begin{equation}
 \bar{\zeta}_{c}=\left.\frac{1}{3}\ln 
\frac{3(\chi_{a}-\sin\chi_{a}\cos\chi_{a})}{2\sin^{3}\chi_{a}}\right|_{\chi_{a}=\pi\sqrt{w}/(1+3w)}.
\label{eq:zetac_chia}
\end{equation}
For $w\ll 1$, this implies
$
 \bar{\zeta}_{c}\approx \pi^{2} w/10.
$
Since $\bar{\zeta}_{c}$ is a monotonically increasing
function of $\chi_{a}$, $\bar{\zeta}_{c}$ takes a maximum value 
$
 \bar{\zeta}_{c}\simeq 0.08602
$
at $w=1/3$ and 
$
 \bar{\zeta}_{c}=(1/3)\ln[(\pi/2-1)/\sqrt{2}]\simeq 0.06377
$
at $w=1$.
As for the peak value $\zeta$, we use the approximate expression (\ref{eq:zeta_chia}).
Then, for type I fluctuations, we find 
$0<\zeta<\zeta_{h}=\ln 2\simeq 0.6931,
$
where $\zeta_{h}$ is the value for $\chi_{a}=\pi/2$.
Our analytic formula gives an expression for
$\zeta_{c}$ as follows:
\begin{equation}
 \zeta_{c}=-2\ln \cos \frac{\pi\sqrt{w}}{2(1+3w)}.
\end{equation}
For $w\ll 1$, this implies
$
 \zeta_{c}\approx \pi^{2}w/4.
$
$\zeta_{c}$ takes a maximum value 
$
 \zeta_{c}\simeq 0.2131
$
at $w=1/3$ and 
$
 \zeta_{c}=-2\ln\cos(\pi/8)\simeq 0.1583
$
at $w=1$.

\begin{center}
\begin{figure}[htbp]
 \includegraphics[width=0.8\textwidth]{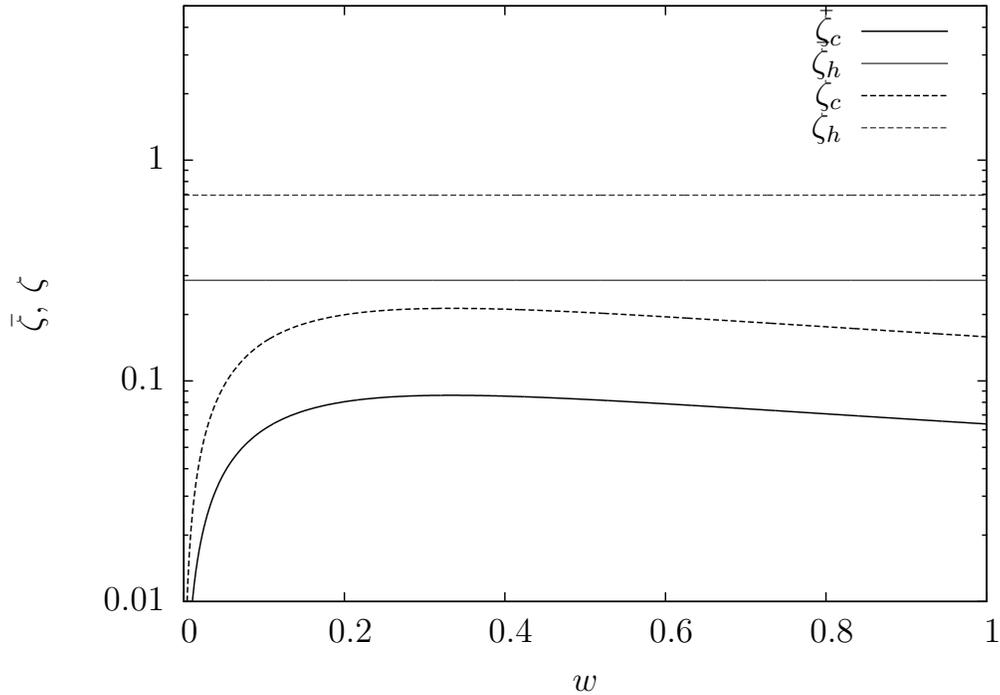}
\caption{\label{fg:curvature_w} 
The threshold values of the curvature perturbations 
$\bar{\zeta}$ and $\zeta$ for different values of $w$. 
The lower thick and upper thin solid lines denote our analytic formula for the threshold
 $\bar{\zeta}_{c}$ and the value $\bar{\zeta}_{h}$ for a
three-hemisphere, respectively. 
The lower thick and upper thin dashed lines denote our analytic formula for the threshold
 $\zeta_{c}$ and the value $\zeta_{h}$ for a
three-hemisphere, respectively, under the approximation described in the text.
The regions below and above the three-hemisphere line 
 correspond to type I and II fluctuations, respectively, for each of
$\bar{\zeta}$ and $\zeta$.
} 
\end{figure}
\end{center}

Figure~\ref{fg:curvature_w} plots our analytic formula for the threshold 
values $\bar{\zeta}_{c}$ and $\zeta_{c}$. For $1/3\alt w\alt 1$, we find 
$0.064\alt \bar{\zeta}_{c}\alt 0.086$ and $0.16\alt \zeta_{c}\alt 0.21$, 
and both are insensitive to $w$.

\section{Summary}

We have introduced an analytic three-zone model to describe primordial 
black hole formation. We then applied this model 
and derived a matter-independent maximum amplitude of 
density perturbation at the horizon crossing time.
Next, we applied the same model to 
the perfect fluid with the equation of state $p=w\rho c^{2}$. 
We then analytically derived a threshold value $\delta_{H c}^{\rm UH}$ 
for the density perturbation 
at the horizon crossing in the uniform Hubble slice 
by a physical argument about the sound waves and 
the maximum expansion. 
We clarified the relationship of the density perturbations 
between the uniform Hubble slice and the comoving slice.
Then, we compared the analytic formula to the result of the 
state-of-the-art numerical simulations 
from the initial data constructed by the first-order asymptotic 
quasihomogeneous solutions.
We have seen that our analytic formula shows a very good agreement 
with the result of the numerical simulations
and the agreement is generally much better than Carr's formula obtained
almost forty years ago.
Further analytic and
numerical studies on this problem will be extremely important to determine the 
threshold and the probability of primordial black hole formation
and then give the precise prediction for the abundance of primordial 
black holes for given early Universe scenarios.

\begin{acknowledgements}
The authors would like to thank C.~T.~Byrnes, 
B.~J.~Carr, T.~Houri, Tsutomu~Kobayashi, 
H.~Maeda, J.~C.~Miller, I.~Musco, T.~Nakama, M.~Sasaki, and T.~Suyama 
for helpful discussions. The authors were partially supported by 
the Grant-in-Aid No. 23654082 (T.H.), and Grants No. 21111006, No. 
22244030, and No. 23540327 (K.K.) for Scientific
Research Fund of the Ministry of Education, Culture, Sports, Science, and
 Technology, Japan. T.H. was also supported by 
Rikkyo University Special Fund for Research.

\end{acknowledgements}

\end{document}